\documentstyle[psfig]{mn}

\def\ra#1#2{#1$^{\rm h}$#2$^{\rm m}$}
\def\dec#1#2{#1$^\circ$#2'}
\def\raa#1#2#3{#1$^{\rm h}$#2$^{\rm m}$#3$^{\rm s}$}
\def\decc#1#2#3{$#1^\circ#2'#3''$}
\def\kms{km~${\rm s}^{-1}$}
\def\msun{M$_{\odot}$}
\def\lsun{L$_{\odot}$}
\def\ha{H$\alpha$}
\def\sii{[S{\sc ii}]}
\def\oi{[O{\sc i}]}
\def\oii{[O{\sc ii}]}
\def\oiii{[O{\sc iii}]}
\def\nii{[N{\sc ii}]}
\def\hasii{H$\alpha$+[S{\sc ii}]}

\begin{document}

\title[New HH flows in Orion]
  {New Herbig--Haro Objects and Giant Outflows in Orion}

\author[S.L. Mader et al.]
{S.L.~Mader,$^1$ W.J.~Zealey,$^1$ Q.A.~Parker,$^{2,3}$ and
M.R.W.~Masheder$^{4,5}$\\
{\small $^1$Department of Engineering Physics, University of Wollongong,
Northfields Avenue, Wollongong NSW 2522, Australia}\\
{\small $^2$Anglo--Australian Observatory, Coonabarabran, NSW 2357,
Australia}\\
{\small $^3$Royal Observatory, Blackford Hill, Edinburgh EH9 3HJ, United
Kingdom}\\
{\small $^4$Department of Physics, University of Bristol BS8 1TL, United
Kingdom}\\
{\small $^5$Netherlands Foundation for Research in Astronomy,
P.O. Box 2, 7990 AA Dwingeloo, The Netherlands}}

\maketitle

\begin{abstract}
We present the results of a photographic and CCD imaging
survey for Herbig--Haro (HH) objects in the L1630 and
L1641 giant molecular clouds in Orion. The new HH flows were
initially identified from a deep \ha\ film from the recently
commissioned AAO/UKST \ha\ Survey of the southern sky. Our
scanned \ha\ and broad band R images highlight both the improved
resolution of the \ha\ survey and the excellent contrast of the \ha\
flux with respect to the broad band R. Comparative IVN survey images
allow us to distinguish between emission and reflection nebulosity.
Our CCD \ha, \sii, continuum and I band images confirm the presence of
a parsec--scale HH flow associated with the Ori I--2 cometary globule
and several parsec--scale strings of HH emission centred on the L1641--N
infrared cluster. Several smaller outflows display one--sided
jets. Our results indicate that for declinations south of -6\degr\
in L1641, parsec--scale flows appear to be the major force in the
large--scale movement of optical dust and molecular gas.
\end{abstract}


\section{Introduction}

The process of star formation is a highly disruptive event where both
infall and outflow of material occur simultaneously in the production of a
protostellar core. The outflow phase is characterised by the impact of
high velocity winds with the surrounding interstellar medium which manifest
as bipolar molecular outflows, Herbig--Haro (HH) objects and jets. 
Multi--wavelength observations have shown HH objects and jets to be
regions of shock--excited gas emitting \ha\ ($\lambda$6563),
\oi\ ($\lambda$$\lambda$6300,6363) and \sii\
($\lambda$$\lambda$6716,6731) in the visible and H$_2$ (2.12$\mu$m)
in the infrared. Their energy sources range from deeply embedded
protostars to optical T--Tauri and Herbig Ae/Be stars.

With the introduction of large format CCD detectors, wide--field
imaging has shown HH flows are more abundant and an order
of magnitude larger than previously thought. Recent narrow band
imaging of the NGC\,1333 star--forming region (SFR) by Bally, Devine
\& Reipurth (1996) found a high concentration of HH objects within a
1/4 square degree region. A similar result was found by Yu, Bally
\& Devine (1997) who conducted a near-infrared H$_2$ (2.12$\mu$m)
survey of the OMC--2 and OMC--3 regions in Orion. Based on
well--studied flows such as HH\,1/2, HH\,34 and HH\,46/47, it was
generally thought their extent ($\sim$ 0.3\,pc) was typical
of outflows from low--mass stars. Bally \& Devine (1994) were the
first to question this view with their suggestion that the HH\,34
flow in Orion is actually 3\,pc in extent. Their idea was confirmed
with deep CCD imaging and proper motion studies of individual knots
(Devine et al. 1997). To date, around 20 giant ($>$1\,pc) HH flows
have been associated with low--mass stars (Eisl\"offel \& Mundt 1997;
Reipurth et al. 1997).

A large number of giant HH flows (and their small--scale counterparts)
may have dramatic effects on the stability and chemical composition of
a giant molecular cloud (GMC). It has been suggested that outflows may
provide a mechanism for self--regulated star--formation and
large--scale bulk motions within GMCs (Foster \& Boss 1996). It is
therefore important to gain information on the
distribution of outflows and particularly giant flows within SFRs.
The new Anglo--Australian Observatory (AAO) and United Kingdom Schmidt
Telescope (UKST) \ha\ Survey of the Southern Galactic Plane
(Parker \& Phillipps 1998a) will be beneficial for such studies as it
provides an unbiased search for new HH objects over entire SFRs with
its wide--field and high resolution capabilities. 

In this paper we concentrate on a search for new HH objects in the
first, deep \ha\ film of the Orion SFR. The distance to the Orion
region lies between 320 to 500\,pc (Brown, De
Geus \& De Zeeuw 1994). Here we adopt a distance of 470\,pc based on
known HH objects in the region (Reipurth 1999). Strong emission and
reflection nebulosity in the region makes searching for HH objects
difficult. Previous attempts at surveys for faint red nebulosities
in L1630 and L1641 have used standard broad band IIIaF R plates
(IIIaF emulsion and RG630 filter), which were limited to subregions
clear of high background emission (Reipurth 1985; Malin, Ogura \&
Walsh 1987; Reipurth \& Graham 1988; Ogura \& Walsh 1991). The new,
deep fine resolution \ha\ films enable us to conduct a more
complete survey for emission--line nebulosities for consequent
follow--up observations.

In Section 2 we present a brief introduction to the specifics
of the \ha\ survey and details on observations and data reduction.
Results are presented in Section 3 where individual objects are
discussed. In Section 4 we make some general conclusions and references
to future work.

\section{Observations and Data reduction}

\subsection{The AAO/UKST \ha\ survey}

Under the auspices of the AAO, the UKST has recently embarked on a
new \ha\ survey of the Southern Galactic Plane, Magellanic Clouds
and selected regions. No systematic high resolution \ha\ survey has
been carried out in the southern hemisphere since the pioneering
work of Gum (1955) and Rodgers, Campbell \& Whiteoak (1963). With
the increase in resolution and sensitivity of differing wavelength
technologies, there has been the need to perform an \ha\ survey
with similar attributes.

The unusually large, single--element \ha\
interference filter is centred on 6590\AA\ with a bandpass of 70\AA. It
is probably the largest filter of its type in use in astronomy. Coated
onto a full field 356mm $\times$ 356mm RG610 glass substrate, the 305mm
clear circular aperture provides a 5.5\degr\ field--of--view.
Further details of the filter properties and specifications are
given by Parker \& Bland--Hawthorn (1998). The detector is the fine
grained, high resolution Tech Pan film which has been the emulsion of
choice at the UKST for the last 4 years. This is due to its excellent
imaging, low noise and high DQE (e.g. Parker, Phillipps \& Morgan 1995;
Parker et al. 1998). Tech Pan also has a useful sensitivity peak at
\ha\ as it was originally developed for solar patrol work. Though
electronic devices such as CCDs are the preferred detector in much of
modern astronomy, they cannot yet match the fine resolution and
wide--field coverage of the Tech Pan film and UKST combination.

Typical deep \ha\ exposures are of 3 hours duration, a compromise
between depth, image quality and survey progress as the films are
still not sky--limited after this time. The Southern Galactic Plane
survey requires 233 fields on 4 degree centres and will take 3 years
to complete. Initial survey test exposures have demonstrated that the
combination of high quality interference filter and Tech Pan film
are far superior for the detection and resolution of faint emission
features near the sky background than any previous combination of
filter and photographic plate used for narrow band observations
(Parker \& Phillipps 1998a). It is the intention that the original
films will be digitised using the Royal Observatory Edinburgh's
SuperCOSMOS facility (Miller et al. 1992). It is planned to release a
calibrated atlas of digital data to the wider astronomical community
as soon as possible.

\subsection{Photographic astrometry and image reduction}

For the Orion region, a deep 3--hour \ha\ exposure was obtained on
1997 December 2nd during a period of good seeing. The plate (HA~17828)
was centred at \ra{05}{36}, \dec{-04}{00} (1950) and designated
grade A based on standard UKST visual quality control procedures by
resident UKST staff. Three independent visual
scans of the film were carefully made by QAP, SLM and WJZ using an
eyepiece and later a 10$\times$ binocular microscope. HH objects
display a wide range of morphologies including knots, arcs and jets.
A combined list of such features was produced and served
as the basis for subsequent astrometry. The new \ha\ images were
then compared with deep non--survey UKST IIIaJ, IIIaF and IVN
broad band copy plates of the same field to confirm the objects as
true emission--line sources. The plates used and their
characteristics are presented in Table \ref{plates}.

Crude positions for each object were first determined using simple XY
positions from the film and transformed to B1950 coordinates by use
of the UKST program PLADAT. Accurate positions were then obtained by
using {\it SkyView} FITS files of the surrounding region. This resulted 
in a positional accuracy within 2\arcsec\ for each object. Digitised
images of each source were then made using a video digitising system
(Zealey \& Mader 1997; 1998). This enabled us to process images via
un--sharp masking and histogram enhancement to recover the original
detail as seen on the TechPan film.

\subsection{CCD observations}

\subsubsection{Optical}

As the Orion region shows highly structured background emission, it is
important we distinguish between photo--ionised filamentary structures
and bona fide HH objects. This can be accomplished with \ha\ and \sii\
images by noting that HH objects usually have \sii/\ha\ ratios $>$ 1
compared to \sii/\ha\ $<$ 1 for emission associated with H{\sc ii}
regions. We obtained narrow and broad band images of HH candidates at
the Australian National University 1.0m telescope at Siding Spring
Observatory during various periods in January--April 1998.
Imaging was done with a 2048 $\times$ 2048 TEK CCD mounted at the
f/8 Cassegrain focus. The 0\farcs6 per pixel gave a field--of--view of
20\farcm48 $\times$ 20\farcm48. The seeing conditions during
usable time was typically $<$ 3\arcsec. Narrow band filters used were
\oiii\ ($\lambda$5016; $\Delta\lambda$ 25\AA), \ha\ ($\lambda$6565;
$\Delta\lambda$ 15\AA), \sii\ ($\lambda$6732; $\Delta\lambda$ 25\AA)
and red continuum ($\lambda$6676; $\Delta\lambda$ 55\AA). The \ha\
filter also transmits the \nii\ ($\lambda\lambda$6548/6584) lines.
We used a standard Kron--Cousins filter for the I band observations.

Typical exposure times were 300s and 900s for broad and narrow band
frames respectively. Flat fields were obtained by illuminating
the dome with a halogen lamp. All frames were reduced in a
similar fashion with IRAF\footnote{IRAF is distributed by the
National Optical Astronomy Observatories, which are operated by
the Association of Universities for Research in Astronomy, Inc.,
under cooperative agreement with the National Science Foundation.},
where 25 median combined bias frames were subtracted from source
frames prior to flat fielding. Individual source frames were median
combined to produce the final images. In several instances, we were
not able to obtain corresponding continuum frames to our CCD \ha\
and \sii\ images. As major HH emission--lines do not fall within
the spectral response curve of the RG715\,+\,IVN filter/emulsion
combination ($\Delta\lambda$ = 6900\AA--9100\AA), we use photographic
IVN images to serve as continuum images where needed.

\subsubsection{Near--infrared}

In January 1993 several HH complexes (including Ori I--2) were imaged using
IRIS, the AAO infrared camera and low resolution spectrograph.
The 128 $\times$ 128 format array has 60$\mu$m pixels which when used in
the f/15 imaging mode provided a spatial resolution of 1\farcs94 per pixel
and a 4\farcm1 $\times$ 4\farcm1 field--of--view. Each source was observed
through a 1\% bandpass filter centred on the H$_2$ $v=1-0~S(1)$ transition
at 2.12$\mu$m. Continuum images were made using a 4\% bandpass filter
at 2.24$\mu$m. Individual frames were linearised, flat fielded against
a dome flat and sky subtracted before being combined and calibrated using
the IRIS image reduction package known as YOGI-FIGARO. A mosaic of twelve
frames, each of five minutes in length were combined to form the final
images.

\section{Results}

In Table \ref{coords} we list new HH objects identified by
our narrow band CCD imaging of candidates identified from the
Orion \ha\ plate. Several of the new objects were identified
by Reipurth (1985) as candidate HH objects from his ESO R film
survey of the Orion region. Objects independently discovered
by the CCD imaging of Reipurth, Bally \& Devine (1998; hereafter
R98) are indicated. In addition to brief comments about their
nature  and location, Table \ref{coords} also suggests possible
energy sources based on evidence presented.

\subsection{New objects in L1630}

Our survey region of the southern portion of L1630 is shown in
Fig. \ref{orionb}. The rest of the cloud complex extends several
degrees to the north--east of the figure. A diffuse shell of
\ha\ emission and a network of bright--rimmed cometary globules
surrounds the multiple OB system $\sigma$ Ori. Ogura
\& Sugitani (1998) list many of these globules as remnant clouds
which may be sites of retarded star formation. The bright
H{\sc ii} regions NGC\,2024 and IC\,434 (which includes The Horsehead
Nebula) outlines an ionisation front between the southern portion of
L1630 and $\sigma$ Ori. This ionisation front extends towards
the open cluster NGC\,1981 which approximately marks the division between
L1630 and L1641. The position of the new HH flows in the region are
indicated.

\subsubsection{HH\,289 (Figs \ref{orionb}--\ref{oi2ccd})}

Located on the north--western outskirts of IC\,434, the bright rimmed
cometary globule Ori I--2 is host to the low--luminosity (L$_{bol}$ =
13 \lsun) IRAS source 05355-0416, which drives both a bipolar CO and
near--infrared molecular hydrogen outflow (Sugitani et al. 1989;
Cernicharo et al. 1992; Hodapp 1994). The IRAS source is also associated
with a H$_2$O maser (Wouterloot \& Walmsley 1986; Codella et al. 1995).

A comparison between our scanned \ha\ and IVN images (Figs \ref{oi2opt}a,b)
identifies a chain of emission--line objects (objects 2--5)
extending to the east of the globule. To the west, we see another
emission--line feature (object 1) which appears as an extension of
faint emission seen in the IVN image. The \hasii\ images (Figs
\ref{oi2ccd}a,b) confirms the presence of a HH flow, designated here as
HH\,289. In the central part of the globule, our \hasii\ and H$_2$ images
(Figs \ref{oi2ccd}b,c) show two faint \sii\ knots (HH\,289 A/B) which
mirror the position of the H$_2$ emission. With the exception of knot
C, all knots appear \sii--bright. Knots D--F show large arc--like
morphologies which open towards the IRAS source. This gives the
impression of a bubble surrounding the eastern side of the globule
which may represent an interface between the outflow and the UV
radiation field from $\zeta$ Ori, which is 42\arcmin\ to the east
of Ori I--2.

From the distribution of optical and near--infrared emission about
the IRAS source (Figs \ref{oi2ccd}b,c), we suggest it is the driving source
of the HH\,289 outflow. The chain extends 551\arcsec\ from the IRAS source
making the lobe 1.23\,pc in projection. This puts the Ori I--2 flow
in the class of parsec--scale flows from low--mass stars (Reipurth, Bally
\& Devine 1997). As HH objects typically
display tangential velocities in the order of 150 \kms\ (i.e., Mundt 1988),
the age span of the optical knots ranges from 530 yr (knot A), to 8100 yr
(knot F). The projected lengths of the (redshifted) CO, H$_2$ and optical
HH flows are 40\arcsec\ (0.09\,pc), 80\arcsec\ (0.18\,pc) and 551\arcsec\
(1.23\,pc) respectively. Apart from knot A, we do not see any evidence of
HH\,emission associated with the blueshifted CO lobe, which we
expect will be very faint due to the tenuous medium on the
western side of the globule. Deeper \sii, \oii\ and/or \oiii\
images of the western side of the globule may reveal fainter
emission.

In Fig. \ref{oi2ccd}b, we note the appearance of a tube--like
feature extending out of the western side of the globule (object
1 in Fig. \ref{oi2opt}a). It is well aligned and mirrors the inner \sii\
and H$_2$ knots with respect to the IRAS source. As this feature
is visible on our Schmidt images, the emission is most probably
scattered light reflected off the walls of the cavity formed
by the outflow as it bores its way out of the globule. Using
AAO/UKST \ha\ material, we have identified a similar feature
associated with the cometary globule and outflow complex CG30/HH\,120
(Zealey et al. 1999). The \ha\ streamer extends to the south--west
of the globule and appears to be the optical counterpart of
an extensive H$_2$ filament associated with the infrared
source CG30--IRS1. The tube--like feature in Ori I--2 and the
streamer in CG30 may represent limb--brightened cavities.

\subsubsection{HH\,444 (Figs \ref{orionb} \& \ref{v510ccd})}

Located in the vicinity of $\sigma$ Ori (Fig. \ref{orionb}),
V510 Ori (= HBC 177) was first classified as a T--Tauri star based
on an objective--prism survey of the Orion region by Sanduleak (1971).
Cohen \& Kuhi (1979) list the star as a classical T--Tauri star (cTTs)
with W(\ha) $>$ 10\AA. The \ha\ emission--line survey of Wiramihardja
et al. (1991) found the source to be a strong \ha\ emitter
with {\sl V} = 14.6 mag as opposed to {\sl V} = 13.54 mag found
by Mundt \& Bastian (1980).

By use of \ha\ material, the first optical detection of the V510 Ori
jet (Parker \& Phillipps 1998b; this paper) is shown in Fig. \ref{v510ccd}a.
The jet has previously been identified by long--slit spectroscopic
studies (Jankovics, Appenzeller \& Krautter 1983; Hirth, Mundt \&
Solf 1997). The scanned \ha\ image (Fig. \ref{beori}a) reveals a highly
collimated jet. Several faint knots (A--C)
are located 57\arcsec, 84\arcsec and 194\arcsec\ from V510 Ori.
The flow terminates at the large bow shock structure HH\,444D,
which displays wide wings which sweep back towards to position
of V510 Ori.

The \hasii\ image (Fig. \ref{v510ccd}b) clearly identifies the
HH\,444 jet extending from V510 Ori. Due to the seeing conditions
at the time ($\sim$ 3\arcsec), we can only confirm the presence
of knots B and D in the \hasii\ image. For the continuum frame
(Fig. \ref{v510ccd}c), conditions were slightly better and based
on the scanned \ha\ and continuum image, knots A--C are considered
as pure emission--line features. The jet appears as two separate parts,
with the first section appearing as a dense region extending 10\arcsec\
from V510 Ori, while a second, more fainter part extends a further
6\arcsec. This change may represent several individual condensations
not resolved by our images. The total projected length
of the optical flow is 0.6\,pc in length.

The small separation between V510 Ori and its jet implies the
jet is still active today and coupled with the fact that we do not
see an obvious counter flow suggests an evolved
case of a one--sided jet (Rodr\'iguez \& Reipurth 1994). High
resolution optical and near--infrared studies of the jet
and energy source will be beneficial in determining the nature of
this unusual outflow complex.

\subsection{New objects in L1641}

As shown in Fig. \ref{oriona}, the northern border of L1641 is
approximated by the bright ionisation front near the open cluster
NGC\,1981. The cloud extends several degrees south of
the figure. The \ha\ emission surrounding the bright H{\sc ii}
region M42 shows remarkable substructure. The southern portion
of the image is bounded by the bright reflection nebulosity
NGC\,1999. In contrast to the L1630 region, we have identified
15 HH\,complexes within the outlined region shown in Fig.
\ref{oriona}. The region is shown in more detail in Fig.
\ref{strings}, where the new objects and features of note are
indicated. Several strings of objects appear to extend to the
north and north--east of the figure. The outlined region towards
the centre of Fig. \ref{strings} contains a cluster of objects
surrounding the high--luminosity source IRAS 05338--0624 (L$_{bol}$
$\sim$ 220 \lsun).

\subsubsection{HH\,292 (Figs \ref{strings} \& \ref{beori})}

Located in the south--east portion of Fig. \ref {strings}, BE Ori
(= HBC 168; IRAS 05345-0635) is a classical T--Tauri star
with W(\ha) $>$ 10\AA\ (Cohen \& Kuhi 1979; Strom, Margulis \& Strom
1989a). No molecular outflow was detected by Levreault (1988). The
near--infrared photometry of Strom et al. (1989a) indicates excess
emission suggesting the presence of a remnant circumstellar disk.

In Fig. \ref{beori}, our scanned \ha\ and CCD images
clearly show a highly collimated flow originating from BE Ori.
The flow has also been identified by Reipurth (1999; private
communication). On the \ha\ scan (Fig. \ref{beori}a), knots
B--D appear to be linked by a stream of \ha\ emission which could
be interpreted as a jet. BE Ori itself is surrounded by diffuse
\ha\ emission which extends towards knot A, which is to the
south--west of the source. All these features are confirmed by
our \hasii\ and continuum images (Figs \ref{beori}b,c). All
knots appear \ha--bright with knot B
displaying a combination of emission and continuum emission.
Designated HH\,292, the flow extends along PA = 45\degr\
with knot A located 114\farcs7 to the south--west and knots B--D
located 21\farcs2, 47\farcs3 and 64\farcs4 to the north--east
of BE Ori respectively, making the total flow length 0.4\,pc.
In their survey of L1641, Stanke, McCaughrean \& Zinnecker
(1998; hereafter SMZ98) identified compact H$_2$ emission
associated with knots A and D (SMZ 25), which may represent
the terminal working surfaces of the flow where the wind is
encountering dense material.

It is interesting to note the asymmetry in HH\,emission with
respect to BE Ori. The lack of optical counterparts to knots
B--D to the south--west of the source suggests BE Ori
has either undergone highly irregular outbursts in the past,
or has a one--sided jet (Rodr\'iguez \& Reipurth 1994). Assuming a
tangential flow velocity of 150 \kms, knots B--D have ages
approximately 300, 700 and 1000 yr respectively, suggesting
periodic outbursts every 300--400 yr whereas knot A has an
age of 1700 yr. As the seeing during our observations of BE Ori was
$\sim$ 3\arcsec, deeper imaging may reveal further HH\,
emission and constrain the ejection history of the source.

\subsubsection*{The L1641--N region}

In Fig. \ref{1641nopt}, we present scanned \ha, IIIaF and IVN images
of the outlined region in Fig. \ref{strings} where a cluster
of faint red nebulous objects was found by Reipurth (1985). The region
has been mapped in $^{12}$CO by Fukui et al. (1986, 1988) who found a
bipolar outflow, L1641--N, centred on the bright far--infrared source
IRAS 05338--0624. Near--infrared imaging of the region by Strom et al.
(1989b), Chen et al. (1993) and Hodapp \& Deane (1993), revealed a dense
cluster of approximately 20 members surrounding the IRAS source.
Davis \& Eisl\"{o}ffel (1995; hereafter DE95) and SMZ98 identified
a multitude of H$_2$ (2.12$\mu$m) emission which outlines a cavity
bored out by the CO outflow and multiple jet and bow shock features
which extend at least 2\,pc to the south of the embedded cluster.

In the following, we present our CCD images of the region shown in
Fig. \ref{1641nopt} which confirm many of the Reipurth nebulosities
as bona fide HH objects. Scanned \ha\ images for several of these
objects are also presented in Parker \& Phillipps (1998b). Independent
CCD imaging of the region has also been presented by R98. Candidate
energy sources for these flows are presented based on their location
with respect to the optical and near--infrared emission (DE95, SMZ98).

\subsubsection{HH\,301/302 (Figs \ref{1641nopt} \& \ref{hh301ccd})}

Extending to the east of Fig. \ref{1641nopt}, the combined \hasii\
image of these two objects (Fig. \ref{hh301ccd}a) shows HH\,301
consists of three bright knots (A--C) which form an U--like structure
with several fainter knots (D--F) trailing to the south--west.
Likewise, HH\,302 consists of one bright knot (A) with a fainter one
(B) extending to the south--west. Both objects are brighter in \sii\
with faint \ha\ emission. This property is apparent from Figs \ref{1641nopt}a
and \ref{1641nopt}b, where HH\,301/302 are prominent on the IIIaF, but
faint in the \ha\ image.  R98 suggest HH\,301/302 are related based on
their elongation towards the L1641--N embedded cluster where the
presumed driving source is located. A line of \sii\ emission
can be seen to the south which mirrors the position of HH\,301/302
and coincides with H$_2$ emission (SMZ 17/18). The bright knot HH\,298A (R98)
can also be seen in Fig. \ref{hh301ccd}a. Although R98 list HH\,298 being
70\arcsec\ in extent with an east--west orientation, our \hasii\
image shows HH\,298 extends even further to the east of HH\,298A
with several knots which we label as HH\,298 D--F. This makes the
HH\,298 flow 340\arcsec, or 0.76\,pc in length from knots A to F.
It is interesting that together with HH\,301/302, HH\,298 produces
a V--type structure with the apex pointing back towards the
infrared cluster.

DE95 and SMZ98 identified a chain of H$_2$ knots (I/J and SMZ 16 A/B
respectively) which extend east from the embedded cluster with a
morphology reminiscent of a jet. In fact, HH\,298A appears directly
between SMZ 16 A and B. As HH\,298 and HH\,301/302 contain both optical
and near--infrared emission, we suggest they are
tracing the walls of a cavity outlined by the V--type structure.
The presence of a jet (SMZ 16A) and counterflow (HH\,298A and SMZ 16B)
suggests we are seeing a single outflow complex. As the jet extends
directly between HH\,298 and HH\,301/302, we do not rule out the
possibility of 3 separate flows, although we draw a comparison with
the outflow source L1551--IRS5, where HH\,28/29 are not located along
the jet axis, but close to the walls of a cavity identified by
optical, near--infrared and CO observations (see Davis et al.
1995 and references therein).

Chen et al. (1993) identified a K\arcmin\ band source (their N23)
in the direction of DE95 I/SMZ 16B which is not visible in our
I band image (Fig. \ref{hh301ccd}c). Based on the alignment of
optical and near--infrared emission, we propose this source as
the driving agent for both HH\,298 and HH\,301/302. Further spectroscopic
studies are needed to clarify its nature.

\subsubsection{HH\,303 (Figs \ref{1641nopt} \&  \ref{hh303ccd})}

The HH\,303 flow consists of two groupings of knots aligned along a
north--south direction. The \hasii\ image in Fig. \ref{hh303ccd}a
shows the northern--most group (knots A-F) outlines a bow--shock with a
sheath of \ha\ emission
overlaying clumpy \sii\ emission. Several more \sii--bright
knots (I--K) extend towards the south. A fainter knot, HH\,298A
(R98) is seen to the south--west of knot K. However, Fig. 5 of
R98 shows HH\,298A at a different location to that shown in
Fig. \ref{hh303ccd}a. Therefore, we identify this knot as HH\,303L in
continuation of R98. R98 suggests HH\,303L may be associated with
HH\,303, but deviates too much from the well defined axis
and may represent a separate flow. We suggest knots I--K and L
represent a remnant bow shock with the former and latter
representing the eastern and western wings respectively.

At first glance, HH\,303 could be interpreted as a highly
collimated flow originating from the variable star V832 Ori
(Fig. \ref{hh303ccd}b). The
optical and near--infrared photometry of this source (source N2
of Chen et al. 1993) shows a spectral energy distribution which
declines rapidly for $\lambda >$ 1$\mu$m, suggesting a lack
of circumstellar material. A comparison of our
optical images with the near--infrared data of SMZ98 shows the
majority of HH\,303 displays both optical and H$_2$
emission, thereby suggesting HH\,303
is behind V832 Ori and unrelated to the star. Knots HH\,
303 B, F and I are coincident with the H$_2$ knots SMZ 8A, 8B,
and 14P respectively, with the H$_2$ emission displaying 
bow shock morphologies which open towards the south in the direction
of L1641--N. 

As knots HH\,303 I--K lie within the blue lobe of
L1641--N, it has been suggested the CO, near--infrared
and optical flows derive from a common source (Strom et al. 1989b;
SMZ98; R98). Chen et al. (1993) identified a bright M band source
(their N15) $\sim$ 8\arcsec\ to the east of the IRAS position.
Chen, Zhao \& Ohashi (1995) detected this source with the VLA at
2.0mm, 7.0mm and 1.3cm, while SMZ98 identified a 10$\mu$m
source coincident with N15 and the VLA source. As N15, the
10$\mu$m source and the 1.3cm source represent the same object,
we follow R98 and label it as the ``VLA source'' which they
suggest is the driving source for HH\,303 and the illuminator
of the reflection nebulosity seen to the north--east in our
I band image (HD93; Fig. \ref{hh303ccd}b).

However, it is important to mention that the L1641--N region is a
highly clustered environment where identifying outflow sources
requires the highest resolution possible. Anglada et al.
(1998) identified two radio continuum sources, VLA2 and VLA3,
which are 0\farcs8 and 0\farcs2 to the west and east
respectively from the nominal position of the VLA source.
Further observations of the region reveal a fainter source
within 1\arcsec\ of VLA2 (Anglada 1998, private communication).
The CO data of Fukui et al. (1986; 1988) clearly indicates the
L1641--N molecular outflow is more complex than a simple bipolar
outflow. Higher resolution studies of these sources are needed
to determine which source is driving the optical and H$_2$ emission.
In particular, it would be interesting to see if the VLA source
displays an elongated radio jet with its long axis pointing in the
direction of HH\,303.

In addition to HH\,303, R98 suggest the VLA source also drives 
HH\,61/62, which are located 46\farcm8 (6.5\,pc) to the south of L1641
(see Fig. \ref{abcd}). If their assumption is correct, the HH\,61/62/303
flow is 7\,pc in length, with the northern lobe only 5\% the length of the
southern lobe. Any shocks associated with the northern lobe will be
extremely faint due to the lack of molecular material as the flow moves
away from L1641.

\subsubsection{HH\,304 (Figs \ref{1641nopt} \&  \ref{hh304ccd})}

Located to north--east of the VLA source, the \sii\ image of HH\,304
(Fig. \ref{hh304ccd}a) shows several compact knots which are
\sii--bright. Knot B is compact with a bow shock structure (knot A)
extending towards the north--east and then curls back to the
north--west. Knots C and D display an opposing bow shock structure,
with knots C and D connected by faint \sii\ emission. The overall
morphology of the system suggests the energy source is located between
knots A/B and C/D. The I band image (Fig. \ref{hh304ccd}b) shows a
compact reflection nebulosity with a tail which mimics part of the
\sii\ emission associated with knots A and B. A reddened source
(which we denote as HH\,304IRS) appears where the reflection emission
is most compact.

The HH\,304 complex is also seen in the H$_2$ mosaic of SMZ98, who
label it SMZ 5. HH\,304A is seen as a bright bar which
extends 6\arcsec\ along an east--west direction. At the
position of the compact reflection nebulosity, a bright H$_2$ knot
is seen, with a trail of H$_2$ emission extending from HH\,304IRS
towards HH\,304C. The appearance of the optical and
near--infrared emission suggests we are seeing two lobes with
knots A and C representing the north--eastern and
south--western working surfaces respectively. HH\,304IRS appears
midway between these two opposing working surfaces.
There are no IRAS or \ha\ emission--line stars at the location of
the reflection nebulosity, which implies a deeply embedded source.

\subsubsection{HH\,305 ( Figs \ref{1641nopt} \& \ref{hh305ccd})}

The HH\,305 outflow appears aligned along a north--south axis centred on
the bright (V $\sim$ 11.3 mag) star PR Ori. With the exception of
knots A and F, all objects are \ha--bright, with knot B displaying an
inverted V--type structure only visible in \ha. Knot A shows a bow shock
structure which opens towards PR Ori. It is interesting to
note that HH\,305E represents the brightest nebulosity in the flow.
The increased brightness could be attributed to the flow encountering
an obstacle of some sort, perhaps in the form of a molecular clump.
The dark lane seen in Figs \ref{strings}, \ref{1641nopt} and \ref{hh305ccd}
represents a change in the molecular distribution in this part of L1641.
At the position of HH\,305E, the flow impacts the molecular cloud and
then deflects to where we
see HH\,305F. Based on their separation from PR Ori, R98 suggest
knots C/D represent an HH\,pair located 16\arcsec\ from
the source. Similarly, knots B/E and A/F represent HH\,pairs located
65\arcsec\ and 108\arcsec\ from PR Ori respectively, making the
total flow length 0.54\,pc.

At present, it is unknown if HH\,305 is being driven by PR Ori or a
more embedded source behind it (R98). In a major
study of {\it Einstein} X--ray sources in L1641, Strom et al. (1990)
identified PR Ori as a low--luminosity (13 \lsun) source with a
spectral type of
K4e$\alpha$ and W(\ha) = 0.5\AA. Their {\sl JHKLM}
photometry indicates a lack of infrared colour excess normally
attributed to a circumstellar disk. Based on their data, PR Ori
appears to be a weak--lined T--Tauri star (wTTs).
Its location with respect to the L1641 molecular cloud
shows it lies in a region of low obscuration and in addition to
the fact that SMZ98 did not detect any H$_2$ emission associated
with HH\,305 rejects the notion of an embedded, more younger
source located behind PR Ori.

If PR Ori is the energy source of HH\,305, it would present
a major discrepancy in star formation theory as wTTs are not thought
to be associated with circumstellar disks and/or outflow phenomenon.
Magazzu \& Martin (1994) identified what was thought to be
a HH flow associated with the wTT, HV Tau. Woitas \&
Leinert (1998) suggested the HH object is actually a companion
T--Tauri star with strong forbidden emission lines whose presence
originally led Magazzu \& Martin to their conclusions. How do we
reconcile the fact that PR Ori is a wTTs with an outflow? The
answer may lie in Table 2 of Strom et al. (1990), who list PR Ori
as an optical double. Our CCD images also show PR Ori as an
extended source, in which case it seems more plausible the
companion (PR Ori--B) is the driving source of HH\,305. Clearly,
further studies of this HH\,complex are needed.

\subsubsection{HH\,306--309 (Figs \ref{strings} \&
\ref{flowopt}--\ref{hh310ccd})}

Figs \ref{strings} and \ref{flowopt} show scanned \ha\ and IVN images
of a string of emission--line objects (HH\,306--309) extending away
from the VLA source and up into the main reflection nebulosity of M42.
A large arcuate structure (HH\,407) can be seen near the bright stars
towards the western border. The large rim of \ha\ emission identified
in Fig. \ref{strings} is seen orientated at PA = 55\degr\ and appears
to surround all objects in the figure. A comparison of the \ha\ and
IVN images confirms all objects are pure emission--line features.

\subsubsection*{The HH flows}

In conjunction with the IVN image (Fig. \ref{flowopt}b), our \hasii\
images confirm all as bona fide HH objects. In Fig. \ref{hh306-8ccd},
the \hasii\ image shows HH\,306 consists of two bright compact knots
(B and F) with a trail of emission extending to the south. A further
knot, HH\,306G, lies to the west which may be unrelated, or part of
an older fragmented shock. HH\,307 consists of several bright knots
which mark the apexes of large arcs or wings which sweep out and open
towards L1641--N. R98 suggest HH\,308 appears as a fragmented
bow shock with knots A and B representing the eastern and
western wings respectively. Located between HH\,308A and B, we note
the presence of a third knot not identified by R98 which we denote
here as HH\,308C. HH\,309 (Fig. \ref{hh309ccd}) shows a similar
structure to HH\,308, with knots A and B representing the first
fragmented bow shock, knot C the second and knots D/E the third.
The reverse bow shock morphology of HH\,309B can be explained
by noting the distribution of \ha\ emission on the scanned \ha\
and CCD \hasii\ images. The knot appears to have curled around the
background emission which may have been responsible to creating
the fragmented appearance of HH\,309.

In searching for further emission north of HH\,309, R98 discovered 
several bow shock structures, designated HH\,310, within the
main nebulosity of M42 (see Fig. \ref{strings}). The objects are
brighter in \sii\ than in \ha, thus discounting the possibility they
might be photo--ionised rims. We have also imaged these structures
and for completeness, present our \ha, \sii\ and continuum images in
Fig. \ref{hh310ccd}. Our \oiii\ frame (not shown) does not detect the
bow shocks associated with HH\,310, thereby suggesting the flow is
moving with a velocity less than 100 \kms. Our \sii\ and continuum
images (Figs \ref{hh310ccd}a,b) identify several bow shock structures
to the north--west of HH\,310 which are \sii--bright and absent in the
continuum frame. Assuming for the moment these features are bona fide
HH objects, their apparent deviation from the axis defined by HH\,310
can be explained if the flow is being redirected by an obstacle, possibly
the long tongue--like feature which extends from the top of the images. An
alternative explanation is that they form part of a separate flow,
perhaps from the L1641--N region. Spectroscopic observations of these
features are needed to determine if they are HH shocks.

\subsubsection*{The embedded counterflow}

To the south of L1641--N, SMZ98 discovered a long chain of
bow shocks. Designated SMZ 23, the chain consists of at least 7 bow shocks
(A--G) which may represent the redshifted counterflow to HH\,306--310
(this paper, R98). From the $^{13}$CO data of Bally et al. (1987), the
integrated moment map (Fig. \ref{13co}) shows evidence of a cavity created
by SMZ 23. What is interesting about this cavity is its size and orientation
with respect to L1641--N, HH\,306--310 and the large cavity
dubbed by R98 as the ``L1641--N chimney'', which they suggest
has been excavated by the repeated passage of
bow shocks associated with HH\,306--310. The location of individual
knots associated with SMZ 23 appears to trace the western wall
of the southern cavity, suggesting the flow impacts with
the cavity wall which produces the observed emission.
We suggest this southern cavity is being excavated by SMZ
23 as the redshifted flow propagates into and away from
L1641--N. The $^{13}$CO velocity structure of the southern cavity
is evident from 5--8 \kms, with the L1641--N molecular core and the 
``L1641--N chimney'' appearing around 8 and 8--11 \kms\ respectively.
This gives further evidence that the southern cavity and the
``L1641--N chimney'' represent expanding red and blueshifted lobes
centred on the L1641--N region. 

Following similar arguments in R98, we find the dimensions of this
southern cavity to be 5\arcmin $\times$ 12\arcmin\ in length, giving
a total area of $\sim$ 1 $\times$ 10$^{37}$ cm$^2$. Assuming the
intensity in the cavity lies within 3--5 K/\kms, the total mass
excavated by the SMZ23 flow is $\sim$~37--62~\msun. In comparison,
R98 find HH\,306--310 has removed $\sim$ 190 \msun\ of
gas from L1641. Apart from obvious errors in estimating the
$^{13}$CO intensity and cavity size, we should point out we have not
taken into account the possibility the southern cavity may have been
formed by the combined action of more than one outflow.

SMZ 23, HH\,306--309 and HH\,310 all display large bow shock
structures which open towards the L1641--N region where the
presumed energy source lies. As mentioned for HH\,303, the
high degree of clustering about the VLA source confuses
identifying specific energy source(s). However the principal
components HH\,306B, HH\,307A, HH\,308C and HH\,309A are located
806\arcsec, 1152\arcsec, 1331\arcsec, 1955\arcsec\ away from the
position of the VLA source. In addition to HH\,310A (2764\arcsec),
the HH\,306--310 lobe is 6.3\,pc in length. As the SMZ 23 flow
appears to extend further south from SMZ23G (Stanke 1999; private
communication), the geometry of HH\,306--310 and SMZ 23 about the
VLA source and VLA2/VLA3 strongly favours at least one of them as
the energy source of the optical and near--infrared emission. Whichever
of these sources is responsible for the observed emission, the combined
length of HH\,306--310 and SMZ23 lobes is 10.5\,pc. High--resolution
radio studies will be beneficial for identifying radio jets and
their orientation with respect to the optical and near--infrared
emission.

\subsubsection*{The southern L1641 region}

In a search for optical counterparts to HH\,306--310, our deep IIIaF
plate of the southern region of L1641 identifies several features
reminiscent of large bow shocks. The IIIaF image of these features
is shown in Fig. \ref{abcd}, where object A appears as a diffuse
feature and object B appears as a bright nebulosity with a long
curve which extends 16\arcmin\ to the north near object A. At first glance,
object B and HH\,61/62 (the counterlobe to HH\,303; R98) appear
to outline the eastern and western wings of a large fragmented
bow shock structure. Objects C and D appear as large arc--like
structures which open to the north and are 3--4\arcmin\ in
extent. As C and D are located well away from the main cloud,
our line--of--sight increases which may suggest they are not
physically associated with L1641. We should also note that
many of the terminal bow shocks associated with parsec--scale
HH flows show substantial substructure which is lacking from
the IIIaF image. In order to resolve the nature of features C
and D, we obtained \ha\ and \sii\ images, but due to variable
cloud cover, we were not able to classify these objects as bona
fide HH objects. Deeper images and/or spectra of objects A--D are 
required to determine if they are photo--ionised regions or HH objects.

\subsubsection{HH\,403--406 (Figs \ref{strings} \&
\ref{403-6opt})}

To the north--east of Fig. \ref{strings}, a second string of
objects extends away from the L1641--N cluster. HH\,403 and HH\,404
are located well clear of the eastern edge of the L1641 molecular cloud.
Although seeing at the time of observing was $>$ 3\arcsec, our
\ha\ and \sii\ CCD images (not shown) did allow us to classify
these features as genuine HH objects. In Fig.\ref{403-6opt}, the
scanned \ha\ and IVN images show HH\,403 consists of a large
number of emission--line knots in addition to a curved (HH\,403G)
and amorphous feature (HH\,403H) to the south--west. The CCD images
of R98 clearly shows HH\,403 as a highly fragmented object which
is very similar in appearance to HH\,262 (L\'opez et al. 1998). A
further 9\arcmin\ to the north--east, HH\,404 displays a sickle--like
structure not too dissimilar from the HH\,47 jet (Heathcote et al.
1996). As these features are \ha--bright, R98 raised the question
as to whether or not HH\,403/404 are bow shocks or bright rims.
However, based on morphological grounds, they suggest HH\,403/404 are
highly fragmented bow shock structures which point back towards
L1641--N where the presumed energy source lies. Our
contrast--enhanced scanned \ha\ image of the region (Fig.
\ref{403-6opt}a) appears to confirm their suspicion as we
see a lack of background \ha\ emission in the direction of
HH\,403/404 which has probably been removed by the action of
the flow as it propagates away from L1641. 

The scanned \ha\ image identifies several large--scale bow
shocks with HH\,403 and HH\,404 at their apexes. R98 do not detect
these features on their CCD images. Originally thought to be
bright rims, comparison of
the \ha\ emission with the $^{13}$CO data of Bally et al. (1987), 
indicates these ``rims'' do not outline the L1641 molecular
cloud, or any other well--defined $^{13}$CO ridge. The first bow
shock is defined by the arc--like object HH\,403G and HH\,404H
representing the eastern and western wings respectively. The
eastern wing trails 7\arcmin\ to the south before
it blends into the background \ha\ emission. The second bow shock
appears as an extended feature similar in appearance to HH\,403G.
The third bow shock only displays the western wing which
extends northward from the second bow to the apex of HH\,404, which
shows a bright arc with faint \ha\ emission which combine to
form an inverted U--type structure.

North--east of HH\,404, a faint object HH\,405 displays \ha\ emission
extending along PA = 45\degr. R98 suggest the emission is reminiscent
of a jet. A further 6\arcmin\ to the north--east, HH\,406 is a large
diffuse object. Are HH\,405 and H\,406 related to HH\,403/404? The IVN image
(Fig. \ref{403-6opt}b) shows a reddened source (denoted HH\,405IRS)
at the position of HH\,405. A reflection nebulosity is also seen nearby.
The position of the nearest IRAS source, 05347-0545, is shown in our
IVN image. It is a 60 and 100$\mu$m source only, indicating it is heavily
obscured and may be related to HH\,405 and/or HH\,406. Based on the
location of HH\,405IRS with respect to HH\,405/406 and the reflection
nebulosity, we suggest this source is the driving agent for HH\,405
and HH\,406 thereby making the flow length 0.78\,pc in extent. Near--infrared
polarimetry and imaging will be useful for determining if HH\,405IRS or
IRAS~05347-0545 is the illuminator of the reflection emission.

Located to the far south--west of L1641--N, R98 noted
HH\,127 mirrors the position of HH\,404 with L1641--N
positioned at the centre (see Fig. \ref{abcd}). Although HH\,127 lies
at an angle of 10\degr\ from the HH\,403/404 and L1641--N axis,
they suggest HH\,403/404 and HH\,127 represent the blue and
redshifted lobes respectively of a 10.6 parsec--scale flow centred
on the VLA source. Given the clustered nature of potential outflow
sources about the VLA source, proper motion studies of HH\,127 and
HH\,403/404 are highly desirable to constrain the location of their
energy source(s).

\subsubsection{HH\,407 (Figs \ref{strings}, \ref{flowopt} \& \ref{hh407ccd})}

Located 28\farcm3 north--west of L1641--N and within
close proximity to HH\,306--310, Figs \ref{strings} and
\ref{flowopt} identify a large, highly fragmented structure
located in the direction of several bright stars. The \hasii\
image (Fig. \ref{hh407ccd}) confirms it as a bona
fide HH object as it emits predominately in \sii. Knots A and B
display bow shock structures with a streamer (knots C/D)
extending to the south--east. In Figs \ref{strings} and \ref{flowopt},
fainter \ha\ emission extends a further 6\arcmin\ to the south--east
of knots C/D.

As the streamers of HH\,407 point towards the L1641--N region,
it seems probable the energy source lies in that direction.
An examination of the H$_2$ data of SMZ98 does not reveal any
emission extended towards HH\,407. After re--examining our \ha\
plate, we noticed the presence of a large loop--like
structure (hereafter loop A) extending out of the reflection
nebulosity NGC\,1999 and in the direction of HH\,407. Comparison
of our scanned \ha, IIIaF and IVN images (Fig. \ref{1999opt})
indicates loop A is a pure emission--line feature. Although
faintly seen on the IIIaF image, the scanned \ha\ image clearly
distinguishes loop A from background emission.

In a recent study of the NGC\,1999 region, Corcoran \& Ray (1995;
hereafter CR95) discovered a second loop (hereafter loop B) of
\ha\ emission extending west of the NGC\,1999 which delineates
a poorly collimated outflow associated with HH\,35 and represents
the counterflow to the redshifted molecular CO outflow discovered
by Levreault (1988). CR95 suggest the Herbig Ae/Be star V380 Ori
(which illuminates NGC\,1999) drives HH\,35, loop B and the molecular
outflow. The presence of loops A and B suggests the presence of
a quadrupole outflow in NGC\,1999. Using similar arguments as CR95,
we suggest loop A delineates an optical outflow which, in conjunction
with HH\,407, represents a 6.2\,pc lobe at PA = -23\degr\ with respect to
V380 Ori.

In a search for optical counterparts to HH\,407, our deep IIIaF plates
do not reveal any clear candidates, although if we assume loop A and
HH\,407 are propagating out and away from L1641, the southern
counterflow may not yet have emerged from the far side of the
molecular cloud. Stanke (1999; private communication) has identified
a large H$_2$ feature to the south of NGC\,1999 which may represent
an embedded counterflow to loop A and HH\,407 (see Fig. \ref{13co}).
HH\,130 is a large bow shock structure located 8\farcm5 south--east of
NGC\,1999 and has been linked
to HH\,1/2 (Ogura \& Walsh 1992) and V380 Ori (Reipurth 1998).
CR95 suggest the energy source of HH\,130 is located to the
north--east of knot H (see Fig. \ref{1999opt}). If HH\,130 and/or
the H$_2$ feature represents the
counterflow to loop A and HH\,407, the outflow axis would be bent by
up to 10\degr. A similar situation is seen in HH\,127/403/404 (R98),
HH\,110/270 (Reipurth, Raga \& Heathcote 1996) and HH\,135/136
(Ogura et al. 1998). Proper motion and spectroscopic studies of
HH\,130, HH\,407 and the H$_2$ feature are needed to determine 
if their motion and radial velocities are directed away from the
V380 Ori region.

Is V380 Ori the driving source of loop A? In addition to
V380 Ori, CR95 found two K band sources, V380 Ori--B
and V380 Ori--C, within NGC\,1999. By means of speckle--interferometry,
Leinert, Richichi \& Hass (1997) identified V380 Ori as a binary
consisting of a Herbig Ae/Be (V380 Ori) and T Tauri star. High
resolution mm--interferometry of NGC\,1999 will help
clarify which source is driving the optical emission associated with
loop A.

As shown in Fig. \ref{13co}, HH\,306--310, HH\,407 and the $\int$--shaped
filament (Bally et al. 1987; Johnstone \& Bally 1999) lie within the
rim of \ha\ emission identified in Figs \ref{strings} and \ref{flowopt}.
Approximated by an ellipse 13\farcm6 $\times$ 4\arcmin\ (3.6 $\times$
0.54\,pc) in size, we suggest the ellipse has formed due to the combined
action of the HH\,306--310 and HH\,407 flows expelling molecular gas
from the main cloud core. The UV radiation from the nearby bright stars
excites the outer edge of the expanding molecular material which
we see as the \ha\ ellipse. Such a large--scale movement of molecular
gas by parsec--scale HH flows has been suggested for HH\,34 and
HH\,306--310 (Bally \& Devine 1994; R98).

\section{Conclusions \& Future Work}

By use of a single AAO/UKST \ha\ film of the Orion region, we have
identified emission--line nebulosities which resemble bow
shocks, jets and extensive alignments of arc--shaped nebulae
indicating possible giant molecular flows. Subsequent narrow and broad
band CCD imaging has confirmed these features as genuine HH\,
objects tracing outflows ranging in size from a fraction of a parsec
to over 6\,pc in length. In addition to the 3\,pc wide \ha\ rim surrounding
HH\,306--310 and HH\,407, the \ha\ loop (loop A) extending out of the
NGC\,1999 reflection nebulosity have not been identified in previous
studies. Although these
features are faintly visible in our IIIaF images, the excellent contrast
of the \ha\ films with respect to IIIaF and published CCD images of
these regions clearly distinguishes these features from background
emission, thereby allowing a thorough investigation of how outflows
from young stars affect the surrounding interstellar medium.
The lack of optical and molecular emission associated with HH\,403/404,
the presence of the \ha\ rim and the identification of large $^{13}$CO
cavities associated with HH\,34 (Bally \& Devine 1994), HH\,306--310
(R98) and the SMZ 23 counterflow (this paper) suggests that, in the
absence of massive star formation, parsec--scale flows are the dominating
factor in disrupting molecular gas in GMCs. They may also be responsible
for the continuation of star formation beyond the current epoch. The
creation of large--scale cavities seen in $^{13}$CO maps (R98; this
paper) may produce highly compressed regions which collapse
to form a new wave of star formation. In order to test this idea, high
resolution sub--millimetre observations in conjunction with near--infrared
H$_2$ (2.12$\mu$m) imaging will identify and determine the distribution
of newly--forming Class~0 protostars with respect to the CO cavities.

Although we have suggested candidate energy sources for many of the
new HH flows, only a few (Ori I--2, BE Ori and V510 Ori) can be
considered as certain. The identification of at least
4 sources within an arcminute of the VLA source warrants subarcsecond
CO mapping of the region to determine which source is driving
the optical and near--infrared emission associated with HH\,306--310,
HH\,403/404, HH\,407 and SMZ 23. Near--infrared spectroscopy
of proposed outflow sources for HH\,298/301/302, HH\,304, HH\,305 and
HH\,405 will be useful in classifying their nature for comparison with
other HH\,energy sources. To varying degrees, the optical sources BE
Ori and V510 Ori exhibit optical variability and multiple--ejection
events (HH objects). The fact these sources still posses highly
collimated, one--sided jets well after they have emerged from their
parental molecular cloud may provide important insights into jet
evolution.
 
In relation to the newly discovered parsec--scale flows, high
resolution spectroscopy and proper motion studies of individual
knots associated with HH\,61/62/303, HH\,306--310, HH\,127/403/404,
HH\,407 and features A--D to the far south of L1641--N will
determine velocities, excitation conditions and confirm
points of origin. 

Due to the success of the Orion \ha\ film, the Carina, Cha I/II,
Sco OB1, $\rho$ Oph, R Cra and CMa OB1 star--forming regions
are to be surveyed in a similar fashion to that presented in this
paper. The majority of these cloud complexes lie within 500\,pc and
maximise the detection of faint, large--scale flows for comparative
studies with the Orion region where we hope to address the following
questions: 

\begin{itemize}
\item What is the nature of the energy source? Parsec--scale flows are
associated with Class~0, Class~I and optically--visible T--Tauri
stars. Is the parsec--scale phenomenon due to inherent properties
of the energy source?
\item How does the flow remain collimated over such large distances?
Does the nature of the surrounding environment have a
collimating effect?
\item To what extent do parsec--scale outflows affect star formation
within molecular clouds? Is there any evidence for self--regulated
star formation?
\end{itemize}

\bigskip

\section*{Acknowledgements}
We thank the staff of the AAO and particularly the UKST for the
teamwork which makes the \ha\ survey possible. Thanks also
go to the Mount Stromlo Time Allocation Committee for the generous
allocation of time on the 40inch telescope. SLM acknowledges John Bally
for the use of the Bell Labs 7m $^{13}$CO data and Thomas
Stanke for supplying his H$_2$ data of the L1641--N region. Thanks also
go to David Malin at the AAO for providing unsharp--mask prints of the
Orion film. SLM acknowledges the support of a DEET scholarship and an
Australian Postgraduate Award. We thank the anonymous referee for
comments and suggestions which strengthened the paper. This research
has made use of
the Simbad database, operated at CDS, Strasbourg, France and the
ESO/SERC Sky Surveys, based on photographic data obtained using
the UKST which is currently operated by the AAO.

\clearpage

\begin{table*}
\begin{minipage}{12cm}
\centering
\caption{Plates used in the current survey.}
\begin{tabular}{lcccccrc} \hline
Plate&$\alpha_{1950}$&$\delta_{1950}$&Date&Emulsion&Filter&Exp~(min)&Grade\\ \hline
&&&&&&&\\ 
R~3816&\ra{05}{32}&\dec{-04}{00}&77--12--11&IIIaF&RG630&90&a\\
J~3828&\ra{05}{32}&\dec{-04}{00}&77--12--16&IIIaJ&GG395&70&a\\
I~3869&\ra{05}{32}&\dec{-04}{00}&78--01--13&IVN&RG715&60&b\\
HA~17828&\ra{05}{36}&\dec{-04}{00}&97--12--02&4415&HA659&180&a\\
R~7462&\ra{05}{36}&\dec{-09}{00}&82--01--26&IIIaF&RG630&90&bI\\
J~8235&\ra{05}{36}&\dec{-09}{00}&82--11--14&IIIaJ&GG395&65&aT\\
I~7406& \ra{05}{36}&\dec{-09}{00}&81--12--19&IVN&RG715&90&bI\\ \hline
\end{tabular}
\label{plates}
\end{minipage}
\end{table*}

\clearpage

\begin{table*}
\centering
\begin{minipage}{16cm}
\caption{New Herbig--Haro flows in Orion.}
\label{coords}
\begin{tabular}{lccccll} \hline
HH\footnote{Independently identified by Reipurth et al. (1998).}
&$\alpha_{1950}$ &$\delta_{1950}$  &Former &Region &Comment &Energy Source\\
& & &Designation\footnote{Reipurth (1985).}&&\\ \hline
&&&&&\\
289A     &\raa{05}{35}{32.4} &\decc{-01}{46}{52} &      & Ori I--2; L1630&large bow shocks  &IRAS 05355-0416\\
292A$^a$ &\raa{05}{34}{30.7} &\decc{-06}{36}{03} &      & L1641; NGC\,1999&one--sided jet?   &BE Ori\\
301$^a$  &\raa{05}{34}{12.6} &\decc{-06}{23}{06} &Rei 43& L1641; cluster &cup--shaped object&N23?\footnote{
Source in list of Chen et al. (1993).}\\
302$^a$  &\raa{05}{34}{22.3} &\decc{-06}{22}{31} &Rei 44& L1641; cluster &bright object     &N23?$^{c}$\\
303B$^a$ &\raa{05}{33}{52.7} &\decc{-06}{21}{24} &Rei 36& L1641; cluster &VLA jet           &VLA source\\
304A$^a$ &\raa{05}{34}{10.6} &\decc{-06}{16}{42} &Rei 41& L1641; cluster &working surfaces  &HH\,304IRS\\
305F$^a$ &\raa{05}{33}{58.7} &\decc{-06}{21}{31} &      & L1641; cluster &symmetric group   &PR Ori B?\\
306B$^a$ &\raa{05}{33}{40.9} &\decc{-06}{10}{33} &Rei 34& L1641; cluster &bow shock         &VLA source\\
307A$^a$ &\raa{05}{33}{39.7} &\decc{-06}{04}{46} &      & L1641; cluster &large bow shocks  &VLA source\\
308A$^a$ &\raa{05}{33}{51.4} &\decc{-06}{04}{42} &      & L1641; cluster &linear features   &VLA source\\
309A$^a$ &\raa{05}{33}{35.9} &\decc{-05}{51}{29} &Rei 33& L1641; cluster &fragmented        &VLA source\\
403$^a$  &\raa{05}{34}{36.9} &\decc{-05}{55}{01} &      & L1641; cluster &fragmented        &L1641--N\\
404$^a$  &\raa{05}{34}{46.6} &\decc{-05}{46}{14} &      & L1641; cluster &sickle--shaped    &L1641--N\\
405$^a$  &\raa{05}{34}{57.7} &\decc{-05}{45}{20} &Rei 46& L1641; NGC\,1980&faint             &HH\,405IRS\\
406$^a$  &\raa{05}{35}{11.7} &\decc{-05}{41}{05} &      & L1641; NGC\,1980&diffuse           &HH\,405IRS\\
407A     &\raa{05}{32}{43.2} &\decc{-06}{00}{32} &      & L1641; NGC\,1980&fragmented        &V380 Ori?\\
444D     &\raa{05}{37}{18.6} &\decc{-02}{31}{52} &      & IC\,434; L1630  &one--sided jet    &V510 Ori\\ \hline
\end{tabular}
\end{minipage}
\end{table*}

\clearpage

\begin{figure}
\caption{The L1630 survey region. The image was derived from
an unsharp--mask print of the \ha\ plate. The upper north--east
of the image shows the IC\,434 and NGC\,2024 H{\sc ii} regions.
Surrounding the $\sigma$ Ori OB system is a diffuse shell of
\ha\ emission in addition to numerous cometary globules and
ionisation fronts which extend from NGC\,2024 to the open cluster
NGC\,1981. The location of new HH flows are indicated by their
numbers. North is up and east is left in all images.}
\label{orionb}
\end{figure}

\begin{figure}
\caption{Scanned (a) \ha\ and (b) IVN images of the HH 289
outflow in the Ori I--2 cometary globule. The chain of
emission--line objects (1--5) are shown with respect to the
embedded IRAS source (indicated by the cross). Note the bubble--like
structure surrounding the eastern side of the globule.}
\label{oi2opt}
\end{figure}

\begin{figure}
\caption{(a) Combined \hasii\ image of the HH 289 outflow showing the
outer knots C--F. The inner region of the globule seen in (b)
\hasii\ and (c) H$_2$ clearly show knots A and B have both optical
and near--infrared emission. The position of the IRAS source is marked
by the circle in the H$_2$ image. The feature marked ``cavity''
may represent a cavity evacuated by the outflow as is propagates out
of the globule.}
\label{oi2ccd}
\end{figure}

\begin{figure}
\caption{(a) Scanned \ha, (b) \hasii\ and (c) continuum images
of the HH\,444 outflow. The \ha\ and \hasii\ images clearly
show the jet from V510 Ori. The large bow shock structure, HH\,444D,
sweeps back towards V510 Ori. Note the absence of a counterjet
the the south--west.}
\label{v510ccd}
\end{figure}

\begin{figure}
\caption{Unsharp--mask \ha\ scan of the L1641 survey region. The
northern extent of L1641 is indicated by the bright ionisation rim
near the open cluster NGC\,1981. The bright H{\sc ii} region M42
(NGC\,1976) is surrounded by highly structured filaments with several
cometary globules seen to the east. The bright reflection nebulosity
NGC\,1999 is seen at the southern edge of the image. The box outlines the
region shown in Fig. \ref{strings} where a large number of HH objects
have been identified.}
\label{oriona}
\end{figure}

\begin{figure}
\caption{Scanned \ha\ image of the L1641--N region outlined
in Fig. \ref{oriona}. The H{\sc ii} region M42 (NGC\,1976)
is seen to the north--west with the reflection nebulosity
NGC\,1999 seen to the south. New objects are numbered and
indicated by a single object to avoid confusion. A large rim
of \ha\ emission appears to surround the HH\,306--309 and HH\,407
group. To the north of HH\,309, R98 identified several bow shocks
(HH\,310) in the main nebulosity of M42. As a scale reference,
the 3\,pc flow HH\,34 is shown with its northern
(HH\,33/40) and southern (HH\,88) terminal working surfaces.
The central source (34 IRS) is indicated by the cross. The
bordered region (see Fig. \ref{1641nopt}) contains a cluster of
objects surrounding the bright IRAS source 05338-0624
(marked as VLA).}
\label{strings}
\end{figure}

\begin{figure}
\caption{(a) Scanned \ha\ image of BE Ori and the HH\,292 outflow.
A stream of \ha\ emission, or jet, links three knots (B--D) to the
north--east, while a further \ha\ knot (A) is seen to the south--west.
(b) CCD \hasii\ and (c) continuum images of HH\,292. HH\,292B
comprises of both emission and continuum emission.}
\label{beori}
\end{figure}

\begin{figure}
\caption{Scanned (a) \ha, (b) IIIaF and (c) IVN images of the
region outlined in Fig. \ref{strings}. Note that many of the
nebulosities on the IIIaF are rather faint in comparison to
their appearance in \ha. The indicated nebulosities are absent
from the IVN image, confirming they are pure emission--line
objects. The emission--line stars PR Ori and V832 Ori are
indicated and the location of the VLA outflow source (IRAS
05338--0624) is indicated by the box.}
\label{1641nopt}
\end{figure}

\begin{figure}
\caption{(a) \hasii\ and (b) I band images of the HH\,298, HH\,301
and HH\,302 outflows identified from Fig. \ref{1641nopt}. The cross
in the I band image marks the location of the presumed energy source
(N23 of Chen et al. 1993) for HH\,298/301/302.}
\label{hh301ccd}
\end{figure}

\begin{figure}
\caption{(a) \hasii\ and (b) I band images of the HH\,303 complex
identified from Fig. \ref{1641nopt}. The circle marks the location of
the VLA source, which R98 propose as the driving source for HH\,303.}
\label{hh303ccd}
\end{figure}

\begin{figure}
\caption{(a) \sii\ and (b) I band images of the HH\,304
outflow located in the north--east of Fig. \ref{1641nopt}.
Knots A/B and C/D represent opposing bow shocks with a
reddened source located at knot B. The candidate energy
source (HH\,304IRS) displays a fan of reflection nebulosity
extending to the north--east.}
\label{hh304ccd}
\end{figure}

\begin{figure}
\caption{\hasii\ image of the HH\,305 flow identified from
Fig. \ref{1641nopt}. Knot E marks the location where the flow
may be deflected by a dense region indicated by the darkened
strip seen to the south of the image. Note that PR Ori appears
slightly extended and is in fact an optical double, where the
companion is the proposed energy source for HH\,305 (see text).}
\label{hh305ccd}
\end{figure}

\begin{figure}
\caption{Scanned (a) \ha\ and (b) IVN images of the northern
objects HH\,306--309 and HH\,407 identified in Fig. \ref{strings}.
The large rim of \ha\ emission is clearly visible and as no
emission is seen in the IVN image, the rim is identified as a
pure emission--line feature.}
\label{flowopt}
\end{figure}

\begin{figure}
\caption{ \hasii\ image of HH\,306--308. In addition to several
bright knots, the HH\,307 bow shock clearly displays larger bow
shock structures which open to the south and have the bright
knots at the apex of the bows. HH\,308 appears as a highly
fragmented bow shock with HH\,308A displaying an elongated
structure (see text for details).}
\label{hh306-8ccd}
\end{figure}

\begin{figure}
\caption{ \hasii\ image of the HH\,309 bow shock. The
reverse morphology of knot B can be explained as the flow
passes over the background \ha\ emission which extends
from the north--east to the south--west of the figure.}
\label{hh309ccd}
\end{figure}

\begin{figure}
\caption{(a) \sii\ and (b) continuum images of the HH\,310 region
(see Fig. \ref{strings} for location). To the north--west we see
several ``bow shocks'' which may be genuine HH objects associated
with HH\,310. They may be deflected from the flow axis by the
tongue--like feature which extends from the top of the figure.}
\label{hh310ccd}
\end{figure}

\clearpage

\begin{figure}
\caption{Integrated $^{13}$CO map of the L1641 cloud from
Bally et al. (1987). The emission has been integrated from 5.4 \kms\
to 11.4 \kms\ with respect to the L1641--N cloud velocity
(8.4 \kms). Filled and open circles represent the HH\,306--310
and SMZ 23 flows respectively. The western wall of the southern cavity
is traced by the SMZ 23 flow. A H$_2$ feature (+) may represent the
embedded counterflow to HH\,407 ($\otimes$). The VLA and V380 Ori
outflow sources are indicated. The location of the \ha\ rim with
respect to HH\,306--310 and HH\,407 is indicated by the ellipse.
The $\int$--shaped filament (Bally et al. 1987) is seen to the north
and approximates the western wall of the ``L1641--N chimney.'' The
wedge shows intensity in units of K/\kms.}
\label{13co}
\end{figure}

\begin{figure}
\caption{IIIaF image of features A, B, C and D to the far south of
L1641--N. To the south--west the cloud boundary is clearly seen with
respect to the background star field. The features are not visible
on IIIaJ and IVN plates which suggests they are emission--line objects,
possibly HH objects originating from the L1641--N region. The position
of the VLA source is marked by a box for reference and comparison
with Fig. \ref{strings}. The near--infrared counterlobe to HH\,306--310,
SMZ 23, is indicated by open circles. HH\,61/62 are thought to be
associated with HH\,303, while HH\,127 represents a possible counterlobe
to HH\,403/404 (R98).}
\label{abcd}
\end{figure}

\begin{figure}
\caption{Scanned (a) \ha\ and (b) IVN images of the HH\,403/404
and HH\,405/406 complexes identified in Fig. \ref{strings}. The
\ha\ image has been enhanced so as to show the HH\,403/404
outflow has cleared away a significant portion of dust in the
region as it propagates away from the L1641--N region. HH\,403/404
are located at the apexes of two large bow shocks (bow 1 and bow 
3 respectively). The IVN image identifies a reddened source, 
HH\,405IRS, which has associated reflection nebulosity and is the
proposed energy source for HH\,405/406. The nearby IRAS
source 05347--0545 may be related to HH\,405/406
and/or the reflection nebulosity between it and HH\,405IRS. The major
and minor axes of the IRAS error ellipse have been multiplied by two
for clarity.}
\label{403-6opt}
\end{figure}

\begin{figure}
\caption{\hasii\ image of the fragmented HH\,407 bow shock identified
in Figs \ref{strings} and \ref{flowopt}. Knots A and B display arcuate
morphologies. Objects C and D are part of a streamer which trails
several arcminutes to the south--east.}
\label{hh407ccd}
\end{figure}

\begin{figure}
\caption{Scanned (a) \ha, (b) IIIaF and (c) IVN images of the
NGC\,1999 region. The \ha\ image clearly identifies two loops of
emission extending out of NGC\,1999. HH\,130 is a large arcuate object
which extends from the bright bow shock HH\,130A to HH\,130H. The
IVN image indicates the positions of V380 Ori and VLA1, which are
the illuminating and driving sources of NGC\,1999 and HH\,1/2
respectively.}
\label{1999opt}
\end{figure}

\end{document}